\documentclass[12pt]{article}
\usepackage{amssymb,amsfonts}
\usepackage{epsf,epsfig}
\def \ep{\varepsilon}
\def \di{\displaystyle}

\begin{document}
\baselineskip 18pt
\title
{Two-magnon Raman scattering in spin-ladders with exact singlet-rung ground state}
\author{P.~N.~Bibikov\thanks{bibikov@PB7855.spb.edu}}
\date{\it V.~A.~Fock Institute of Physics,\\ Sankt-Petersburg State
University, Russia}

\maketitle

\vskip5mm

\begin{abstract}
Using coordinate Bethe ansatze we construct two-magnon states for
the suggested by A. K. Kolezhuk and H.-J. Mikeska family of
spin-ladder models with exact singlet-rung vacuum. The explicit
formula for zero-temperature Raman scattering cross section is
derived. The corresponding line-shapes are strongly asymmetric and
their singularities originate only from bound states. This form of
a line-shape is in a good correspondence to the experimental data.
\end{abstract}

\section{Introduction}
Raman scattering in spin-ladders was studied in a number of papers
(see \cite{1}-\cite{9} and references therein). The obtained
experimental data was analyzed by several theoretical approaches
\cite{3},\cite{7},\cite{8}. However in none of these papers the
exact formula for the Raman cross section was used. In the present
paper we obtain the {\it exact} formula for the special class of
spin-ladder models with exact singlet-rung vacuum. This family of
models was first suggested in \cite{10}. The corresponding
Hamiltonian $\cal H$ has the following form:
\begin{equation}
{\cal H}=\sum_{n=-\infty}^{\infty}H_{n,n+1},
\end{equation}
where
\begin{equation}
H_{n,n+1}=H^{stand}_{n,n+1}+H^{frust}_{n,n+1}+H^{cyc}_{n,n+1}+H^{norm}_{n,n+1},
\end{equation}
and
\begin{eqnarray}
H^{stand}_{n,n+1}&=&\frac{J_{\bot}}{2}({\bf S}_{1,n}\cdot{\bf
S}_{2,n}+{\bf S}_{1,n+1}\cdot{\bf S}_{2,n+1})+ J_{\|}({\bf
S}_{1,n}\cdot{\bf S}_{1,n+1}+{\bf S}_{2,n}\cdot{\bf
S}_{2,n+1}),\nonumber\\
H^{frust}_{n,n+1}&=&J_{frust}({\bf S}_{1,n}\cdot{\bf
S}_{2,n+1}+{\bf S}_{2,n}\cdot{\bf
S}_{1,n+1}),\nonumber\\
H^{cyc}_{n,n+1}&=&J_c(({\bf S}_{1,n}\cdot{\bf S}_{1,n+1})({\bf
S}_{2,n}\cdot{\bf S}_{2,n+1})+({\bf S}_{1,n}\cdot{\bf
S}_{2,n})({\bf S}_{1,n+1}\cdot{\bf S}_{2,n+1})\nonumber\\
&-&({\bf S}_{1,n}\cdot{\bf S}_{2,n+1})({\bf S}_{2,n}\cdot{\bf
S}_{1,n+1})),\nonumber\\
H^{norm}_{n,n+1}&=&J_{norm}I.
\end{eqnarray}
Here ${\bf S}_{i,n}$ ($i=1,2;\,n=-\infty...\infty$) are
spin-$\frac{1}{2}$ operators associated with cites of the ladder
and $I$ is an identity matrix. The auxiliary term
$H_{n,n+1}^{norm}$ in (2) is need only for normalization to zero
the lowest eigenvalue of the $16\times16$ matrix $H$ of rung-rung
interaction.

It was shown in \cite{10} that when the following conditions
\begin{eqnarray}
J_{frust}&=&J_{\|}-\frac{1}{2}J_c,\quad J_{norm}=\frac{3}{4}J_{\bot}-\frac{9}{16}J_c,\nonumber\\
J_{\bot}&>&2J_{\|},\quad J_{\bot}>\frac{5}{2}J_c,\quad
J_{\bot}+J_{||}>\frac{3}{4}J_c.
\end{eqnarray}
are satisfied then the lowest (zero eigenvalue) eigenstate of $H$
is $w\otimes w$, where $w$ is the rung-singlet state. In this case
the ground state of the Hamiltonian (1) has the simple
tensor-product form:
\begin{equation}
|0\rangle=\prod_n\otimes w_n.
\end{equation}

In order to obtain the full spectrum of $H$ we shall also define
the following triplet states:
\begin{equation}
f^k_n=({\bf S}^k_{1,n}-{\bf S}^k_{2,n})w_n,\quad ({\bf
S}_{1,n}^j+{\bf S}_{2,n}^j)f_n^k=i\varepsilon_{jkm}f_n^m.
\end{equation}

All other eigenstates of $H$ are separated into the following
sectors: singlet $f^k\otimes f^k$, triplet
$\varepsilon_{ijk}f^j\otimes f^k$, quintet $t_{ijkl}f^j\otimes
f^k$ with eigenvalues: $\ep_0=J_{\bot}-2J_{\|},\quad
\ep_1=J_{\bot}-J_{\|}-\frac{1}{4}J_c,\quad
\ep_2=J_{\bot}+J_{\|}-\frac{3}{4}J_c$, and two triplets $w\otimes
f^k\pm f^k\otimes w$ with eigenvalues:
$\ep_{\pm}=\frac{1}{2}(J_{\bot}-\frac{3}{2}J_c\pm J_c)$. Here
$t_{ijkl}=\delta_{ik}\delta_{jl}+\delta_{il}\delta_{jk}-
\frac{2}{3}\delta_{ij}\delta_{kl}$.

The Hamiltonian (1)-(3) commutes with the following magnon number
operator ${\cal Q}=\sum_{n}Q_n$, where $Q_n=\frac{3}{4}I+{\bf
S}_{1,n}\cdot{\bf S}_{2,n}$ is the associated with the $n$-th rung
projection operator on triplet states.

\section{The two-magnon states}

Corresponding to (1)-(4) one-magnon states were obtained in
\cite{10}. Suggesting the following Bethe form for two-magnon
states $|S,\beta\rangle$ (where $S$ is the total spin and $\beta$
the list of additional parameters):
\begin{eqnarray}
|0;\beta\rangle&=&\sum_{m=-\infty}^{\infty}\sum_{n=m+1}^{\infty}a_0(m,n;\beta)\ldots
w_{m-1} f_m^jw_{m+1}\ldots w_{n-1}f_n^jw_{n+1}\ldots\,,
\\
|1;\beta\rangle_i&=&\sum_{m=-\infty}^{\infty}\sum_{n=m+1}^{\infty}a_1(m,n;\beta)\varepsilon_{ijk}
\ldots w_{m-1}f_m^jw_{m+1}\ldots w_{n-1}f_n^kw_{n+1}\ldots\,,
\\
|2;\beta\rangle_{ij}&=&\sum_{m=-\infty}^{\infty}\sum_{n=m+1}^{\infty}a_2(m,n;\beta)t_{ijkl}\ldots
w_{m-1} f_m^kw_{m+1}\ldots w_{n-1}f_n^lw_{n+1}\ldots
\end{eqnarray}
we obtain in standard way \cite{11} the following ${\rm Schr\ddot
odinger}$ equation:
\begin{eqnarray}
&\frac{\displaystyle
J_c}{\displaystyle2}{[}a_S(m-1,n;\beta)+a_S(m+1,n;\beta)+a_S(m,n-1;\beta)+a_S(m,n+1;\beta){]}&
\nonumber\\
&+(2J_{\bot}-3J_c)a_S(m,n;\beta)=Ea_S(m,n;\beta),&
\end{eqnarray}
and Bethe condition for amplitudes:
\begin{equation}
2\Delta_Sa_S(n,n+1;\beta)=a_S(n,n;\beta)+a_S(n+1,n+1;\beta).
\end{equation}
Here $\Delta_S=\frac{\di\ep_S-\ep_+-\ep_-}{\di\ep_+-\ep_-}$.

For each $S$ the Eq. (11) has two solutions. The scattering
solution:
\begin{equation}
a_S^{scatt}(m,n;k_1,k_2)=C_{S,12}{\rm
e}^{i(k_1m+k_2n)}-C_{S,21}{\rm e}^{i(k_2m+k_1n)},
\end{equation}
with $C_{S,ab}=\cos\frac{k_a+k_b}{2}-\Delta_S{\rm
e}^{i\frac{k_a-k_b}{2}}$, and the bound solution:
\begin{equation}
a_S^{bound}(m,n;u)={\rm e}^{iu(m+n)+v(m-n)},
\end{equation}
where the real parameters $v\geq0$ and $-\pi<u\leq\pi$ satisfy the
following condition:
\begin{equation}
\cos u=\Delta_S{\rm e}^{-v}.
\end{equation}
From (14) and nonnegativity of $v$ follows that
\begin{equation}
|\cos u|\leq|\Delta_S|\leq {\rm e}^v.
\end{equation}

The corresponding to (12) and (13) eigenvalues are:
\begin{eqnarray}
E_S^{scatt}(k_1,k_2)&=&2J_{\bot}-3J_c+J_c(\cos
k_1+\cos k_2),\\
E_S^{bound}(u)&=&2J_{\bot}+(\Delta_S-3)J_c+\frac{J_c}{\Delta_S}\cos^2u.
\end{eqnarray}

As we see from (12) and (13) the translation invariant states
correspond to $a_S^{scatt}(m,n;k,-k)$, $a_S^{bound}(m,n;0)$ and
$a_S^{bound}(m,n;\pi)$.

\section{Calculation of Raman cross section}
Following Sugai \cite{2} we shall consider only the case when the
incident and scattered light have parallel polarization directions
both lying in the plane of the ladder and forming an angle
$\theta$ with respect to vertical bonds (rungs). The
zero-temperature two-magnon Raman scattering cross section as a
function of frequency and $\theta$ can be expressed using Fermi's
golden rule \cite{3},\cite{4}:
\begin{equation}
I(\omega,\theta)=\lim_{N\rightarrow\infty}\frac{2\pi}{2N+1}\sum_{\alpha}|\langle\alpha|{\cal
H}^R(\theta)|0\rangle|^2 \delta(\omega-E_{\alpha}),
\end{equation}
where $2N+1$ is the number of rungs. Within the
Fleury-Loudon-Elliot approach the effective Raman Hamiltonian
${\cal H}^R(\theta)$ have the following form \cite{1},\cite{5} (we
also take into account interactions across diagonals):
\begin{equation}
{\cal H}^R(\theta)=A_{leg}\cos^2\theta {\cal
H}^{leg}+A_{diag}(\cos^2(\theta+\gamma){\cal
H}^{d_1}+\cos^2(\theta-\gamma){\cal H}^{d_2})+A_{rung}\sin^2\theta
{\cal H}^{rung}.
\end{equation}
Here $A_{leg}$, $A_{diag}$ and $A_{rung}$ are constants and
$\gamma$ is the angle between rung and diagonal directions.
Operators ${\cal H}^{rung}$ ${\cal H}^{leg}$ and ${\cal H}^{d_1}$,
${\cal H}^{d_2}$ are the following:
\begin{equation}
{\cal H}^{rung}=\sum_n{\bf S}_{1,n}\cdot{\bf S}_{2,n},\quad {\cal
H}^{leg}=\sum_{i,n}{\bf S}_{i,n}\cdot{\bf S}_{i,n+1},\quad {\cal
H}^{d_{1(2)}}=\sum_n{\bf S}_{1(2),n}\cdot{\bf S}_{2(1),n+1}.
\end{equation}

Expressing ${\cal H}^{leg}$, ${\cal H}^{d_1}$, ${\cal H}^{d_2}$
and ${\cal H}^{rung}$ from the auxiliary operators:
\begin{equation}
{\cal H}^{\pm\pm}=\sum_n({\bf S}_{1,n}\pm{\bf S}_{2,n})\cdot({\bf
S}_{1,n+1}\pm{\bf S}_{2,n+1}).
\end{equation}
and taking into account the Eq. (6) we represent
$I(\omega,\theta)$ in the factorized form:
\begin{equation}
I(\omega,\theta)=\frac{1}{4}{[}(A_{leg}+A_{diag}\sin^2\gamma)\sin^2\theta+
A_{diag}\cos^2\gamma\cos^2\theta{]}^2I_0(\omega),
\end{equation}
where
\begin{equation}
I_0(\omega)=\lim_{N\rightarrow\infty}\frac{2\pi}{2N+1}\sum_{\alpha}|\langle\alpha|{\cal
H}^{--}|0\rangle|^2 \delta(\omega-E_{\alpha}).
\end{equation}

Formula (22) expresses the polarization angle dependence of Raman
cross section however it may be applied in a straightforward way
only for $\theta=\frac{\displaystyle m\pi}{\displaystyle2}$
\cite{6}.

From the Eq. (6), translational and ${\rm SU}(2)$ invariance of
${\cal H}^{--}$ follows that only translation invariant singlet
two-magnon states contribute to the formula (23). Separating the
contributions from scattering and bound states we obtain:
\begin{eqnarray}
I_0^{scatt}(\omega)&=&\lim_{N\rightarrow\infty}\sum_{k}\frac{|\sum_{n=-N}^Na(n,n+1;k,-k)|^2}
{\sum_{n=-N+1}^N\sum_{m=-N}^{n-1}|a(m,n;k,-k)|^2}\delta(\omega-E_0^{scatt}(k,-k)),\\
I_0^{bound}(\omega)&=&\lim_{N\rightarrow\infty}\sum_{u=0,\pi}\frac{|\sum_{n=-N}^Na(n,n+1;u)|^2}
{\sum_{n=-N+1}^N\sum_{m=-N}^{n-1}|a(m,n;u)|^2}\delta(\omega-E_0^{bound}(u)).
\end{eqnarray}

From (12) and (13) follows:
\begin{eqnarray}
\sum_{n=-N+1}^N\sum_{m=-N}^{n-1}|a_0^{scatt}(m,n;k,-k)|^2&=&4N^2(1-2\Delta_0\cos
k+\Delta_0^2)+O(N),\\
|\sum_{n=-N}^Na^{scatt}(n,n+1;k,-k)|&=&4N\sin k+O(1),\\
\sum_{n=-N+1}^N\sum_{m=-N}^{n-1}|a_0^{bound}(m,n;u)|^2&=&\frac{2N}{e^{2v}-1}+o(N),\quad u=0,\pi\\
|\sum_{n=-N}^Na^{bound}(n,n+1;u)|&=&(2N+1)e^{-v},\quad u=0,\pi.
\end{eqnarray}
Using the substitution
$\sum_k\rightarrow\frac{2N+1}{2\pi}\int_0^{2\pi} dk$ we obtain
from (26)-(29) the final expressions for the cross sections:
\begin{eqnarray}
I_0^{scatt}(\omega)&=&\frac{4\Theta(1-x^2)\sqrt{1-x^2}}
{J_c(1+\Delta_0^2-2x\Delta_0)},\\
I_0^{bound}(\omega)&=&
\frac{2\pi}{J_c}(1-\frac{1}{\Delta_0^2})\Theta(\Delta_0^2-1)\delta(2x-\Delta_0-\frac{1}{\Delta_0}).
\end{eqnarray}
Here $\Theta$ is the step function and
$x=\frac{\omega-2J_{\bot}+3J_c}{2J_c}$ is the rescaling parameter.

From (15) and (31) follows that the contribution from bound states
$I^{bound}_0(\omega)$ exist only for $|\Delta_0|>1$. The behavior
of $I_0^{scatt}$ as a function of $x$ also essentially depends on
the parameter $\Delta_0=\frac{\di3}{\di2}-2\frac{\di J_{\|}}{\di
J_c}$. When $\Delta_0=\pm1$ the formula (30) reduces and the
line-shape has a singularity at $x=\Delta_0$. For $\Delta_0=1$ it
lies in the top of the two-magnon continuum however for
$\Delta_0=-1$ in the bottom. For $\Delta_0\neq\pm1$ the cross
section $I^{scatt}_0$ is a regular function of $x$ and has the
maximum in the point $x_{max}=\frac{2\Delta_0}{\Delta_0^2+1}$.

In order to study the line-shape in more detail we shall find its
inflection points. Calculating the second derivative of
$I^{scatt}_0$ with respect to $x$ we obtain the following
condition:
\begin{equation}
p(x,\Delta_0)=4\Delta_0(1+\Delta_0^2)x^3-12\Delta_0^2x^2-\Delta_0^4+6\Delta_0^2-1=0.
\end{equation}
Since $p(\pm1,\Delta_0)=-(1\mp\Delta_0)^4$, the polynomial
$p(x,\Delta_0)$ for $\Delta_0\neq\pm1$ has only 0 or 2 zeros in
the interval $(-1,1)$. From standard calculation follows that
$p(x,\Delta_0)$ has the maximum
$p_{max}=-\Delta_0^4+6\Delta_0^2-1$ in the point $x=0$. It is
evident now that for $p_{max}>0$ the line-shape of $I^{scatt}_0$
has two inflection points. From the straightforward calculation
follows that $p_{\max}>0$ only for
\begin{equation}
\Delta_-<|\Delta_0|<\Delta_+,
\end{equation}
where
$\Delta_{\pm}=\sqrt{3\pm2\sqrt{2}}$ ($\Delta_-\approx0.4142$,
$\Delta_+\approx2.4142$). It may be easily proved in a
straightforward way that $(\Delta_+-\Delta_-)^2=4$, so
$\Delta_+-\Delta_-=2$.

In the case (33) the line-shape near the $x_{max}$ is similar to
van-Hove singularity. For $\Delta_-<\Delta_0<\Delta_+$ this
"singularity" lies near the top of the two-magnon continuum
however for $-\Delta_+<\Delta_0<-\Delta_-$ near the bottom. In
both the cases the line-shape of Raman scattering is strongly {\it
asymmetric}. The case $p_{max}<0$ with no inflection points may be
interpreted as a broad maximum. Some line shapes corresponding to
different values of $\Delta_0$ are presented in the Fig. 1.

As it follows from (16) and (17) for
$\Delta_0\rightarrow\pm1+0^{\pm}$ the top (bottom) of the
two-magnon continuum and the bound two-magnon state have the same
energy: $2J{\bot}-3J_c\pm2J_c$. It was proposed in \cite{3} that
in this case the resonance between bound and scattering states
leads to a redistribution of Raman intensity and merging of
singularity. However as we see from (30) and (31) in our model
this conjecture fails. Moreover the singularity in $I^{scatt}_0$
appears only in the resonance $\Delta_0=\pm1$ cases.

\section{Comparison with experiment and discussion}
Raman scattering in ${\rm MgV}_2{\rm O}_5$ and ${\rm CaV}_2{\rm
O}_5$ were reported in \cite{9}. It was pointed that for both
materials the corresponding line-shapes are strongly asymmetric
and have one maximum instead of two. This fact was considered as
strange and there were suggested some conjectures to interpret it.
For example it was supposed in \cite{9} that in ${\rm MgV}_2{\rm
O}_5$ there is no spin-gap and the magnetic ordering is 2D, or the
spin-gap is so small (about 10 ${\rm cm}^{-1})$ that can not be
observed by the used experimental resolution. The asymmetry of the
line-shape for ${\rm CaV}_2{\rm O}_5$ was interpreted in \cite{9}
as originating from next-nearest neighbor interactions. In
\cite{7} it was conjectured that the second peak in the line-shape
of ${\rm CaV}_2{\rm O}_5$ is not observed because it is dominated
by phonon peak. In \cite{3} it was conjectured that the asymmetry
originates from resonance with two-triplet bound state.

In our paper we have demonstrated that the line-shape asymmetry in
spin-ladder Raman scattering is not something strange and
outstanding but may appear in a sufficiently big class of models.
Of course we do not pretend that for some values of exchange
parameters our toy model necessary describes the real materials
such as ${\rm CaV}_2{\rm O}_5$ or ${\rm MgV}_2{\rm O}_5$.
Nevertheless perhaps the true ground state is in some sense
"close" to our idealized one (5) and we may believe that our model
correctly represents some general qualitative features of real
materials. In this context we emphasize that the {\it exactly
calculated} Raman scattering line-shape may be strongly asymmetric
without any additional assumptions such as next-nearest neighbor
interactions, resonance with bound state or dominating by phonon
peak.

The author is very grateful for discussions with K.~P.~Schmidt,
P.~P.~Kulish, and L.~V.~Prokhorov.

\begin{center}
\begin{figure}
\epsfig{file=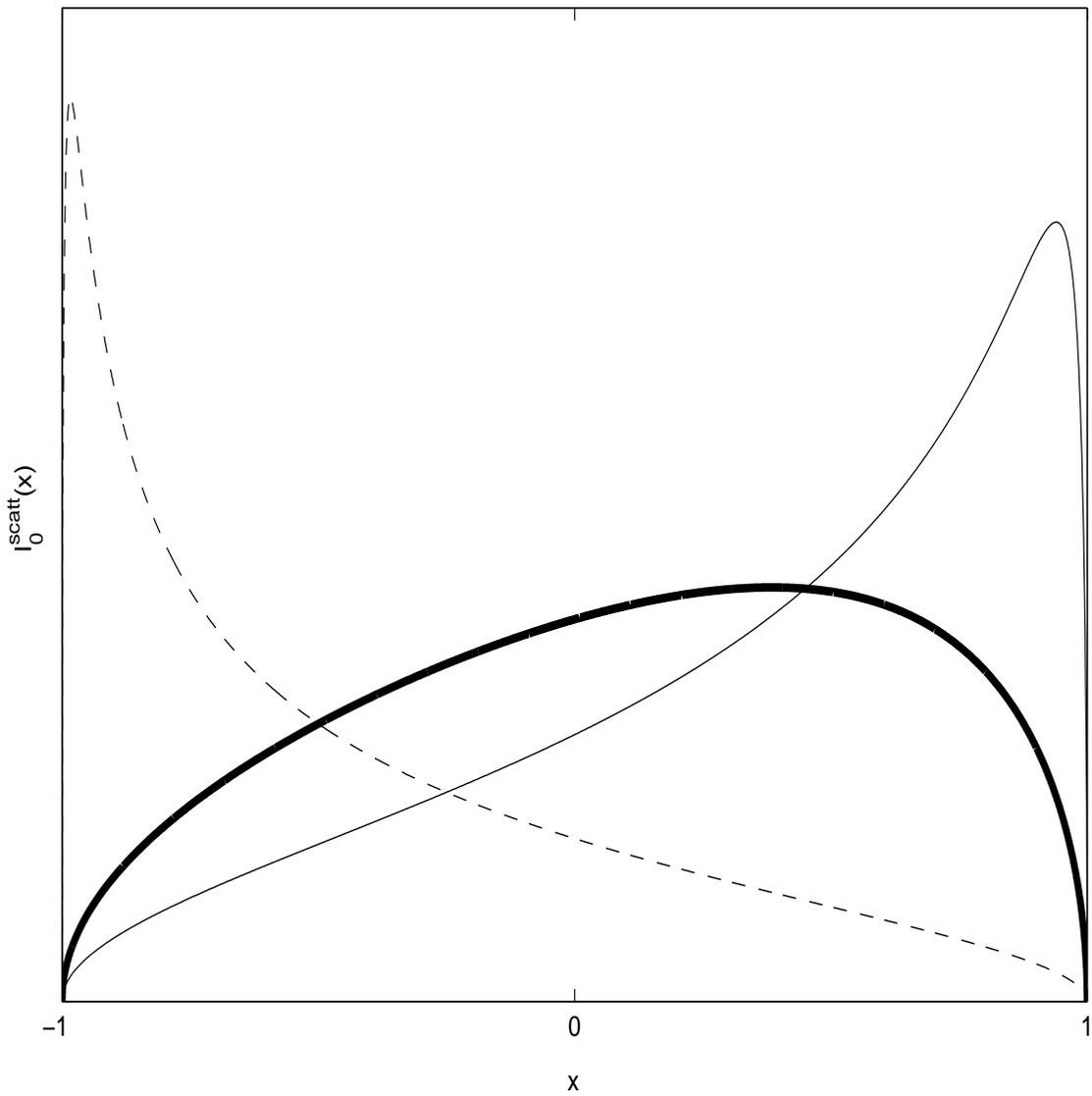, width=15 cm, height=15 cm} \caption {The
thick line: $\Delta_0=0.2$, the thin line: $\Delta_0=0.7$, the
dash line: $\Delta_0=-1.2$. }
\end{figure}
\end{center}


\begin{thebibliography}{11}
\bibitem{1} P.~Lemmens, G.$\rm G{\rm\ddot u}ntherodt$, C. Gros,
Phys. Rep. {\bf 375}, 1 (2003)
\bibitem{2} S.~Sugai and M.~Suzuki, Phys. Status Solidi B {\bf
215}, 653 (1999)
\bibitem{3} C.~Jureska, V.~$\rm Gr{\ddot u}tzun$, A.~Friedrich,
and W.~Brenig, Eur. Phys. J. B {\bf 21}, 469 (2001)
\bibitem{4} E.~Orignac and R.~Citro, Phys. Rev. B {\bf62}, 8622
(2000)
\bibitem{5} P.~J.~Freitas and R.~R.~Singh, Phys. Rev. B {\bf 62},
14113 (2000)
\bibitem{6} A.~Gozar, Phys. Rev. B {\bf 65}, 176403 (2002)
\bibitem{7} K.~P.~Shmidt, C.~Knetter, M.~${\rm Gr}\ddot{\rm u}{\rm
ninger}$ and G.~S.~Uhrig, Europhys. Lett. {\bf 56}, 877 (2001)
\bibitem{8} Y.~Natsume, Y.~Watabe and T.~Suzuki, Journ. Phys. Soc.
Japan {\bf 67}, 3314 (1998)
\bibitem{9} M.~J.~$\rm Konstantinovi{\acute c}$, Z.~V.~$\rm
Popovi{\acute c}$, M.~Isobe and Y.~Ueda, Phys. Rev. B {\bf 61},
15185 (2000)
\bibitem{10} A.~K.~Kolezhuk and H.-J. Mikeska, Int. J. Mod. Phys.
B {\bf 12}, 2325 (1998)
\bibitem{11} R.~Orbach, Phys. Rev. {\bf 112}, 309 (1958)
\end{thebibliography}
\end{document}